%% file: rdid_10062024.tex
\documentclass[bib]{statapress}

\usepackage[crop,newcenter,frame]{pagedims}

\usepackage{sj}

\usepackage{epsfig}

\usepackage{stata}

\usepackage{amsmath}
\usepackage{amsfonts}
\usepackage{bbm}
\usepackage{bm}
\usepackage{booktabs}

\usepackage{shadow}

\usepackage[normalem]{ulem}

\sjsetissue{$vv$}{$ii$}{$mm$}{$yyyy$}

\usepackage{threeparttable} 
\usepackage{tabularx}

\usepackage{mathtools}

\newcommand{\indep}{\rotatebox[origin=c]{90}{$\models$}}

\newtheorem{assumption}{Assumption}
\newtheorem{definition}{Definition}
\newtheorem{proposition}{Proposition}

\begin{document}

\inserttype[st0001]{article}
\author{Ban and K\'edagni}{%
  Kyunghoon Ban\\Rochester Institution of Technology\\Rochester, NY\\ kban@saunders.rit.edu
  \and
  D\'esir\'e K\'edagni\\UNC-Chapel Hill\\Chapel Hill, NC\\dkedagni@unc.edu
}
\title[Robust difference-in-differences]{rdid and rdidstag: Stata commands for robust difference-in-differences}
\maketitle

\begin{abstract}
This article provides a Stata package for the implementation of the robust difference-in-differences (RDID) method developed in \cite{bk2023rdid}. It contains three main commands: \texttt{rdid}, \texttt{rdid\_dy}, \texttt{rdidstag}, which we describe in the introduction and the main text. We illustrate these commands through simulations and empirical examples.

\keywords{\inserttag, \tt{rdid}, \tt{rdid\_dy}, \tt{rdidstag}, robust DID}
\end{abstract}

\section{Introduction}

In this article, we present the \texttt{rdid}, \texttt{rdid\_dy}, \texttt{rdidstag} commands for estimation and inference on robust difference-in-differences (DID) bounds as developed by \cite{bk2023rdid}.
These commands, summarized in Table \ref{tab:commands}, allows one to analyze the average treatment effects on the treated (ATT) under the canonical $2 \times 2$ DID setting where observational data consist of two different groups (a treatment group and a control group) and two time periods (pre-treatment and post-treatment) as well as the staggered adoption design where there are multiple cohorts with different timings in the treatment adoption.

\begin{table}[htbp]
\centering
\caption{Robust DID commands.} \label{tab:commands}
\vspace{2mm}
\renewcommand{\arraystretch}{1.3}
\begin{tabular}{cc}
\hline 
Command & Description \\
\hline
\begin{minipage}{.2\textwidth}
\texttt{rdid}
\end{minipage}
&
\begin{minipage}{.7\textwidth}
\vspace{1mm}
Estimates robust DID bounds on the ATT under the canonical $2 \times 2$ DID setting.
\vspace{1mm}
\end{minipage}\\

\begin{minipage}{.2\textwidth}
\texttt{rdid\_dy}
\end{minipage}
&
\begin{minipage}{.7\textwidth}
\vspace{1mm}
Estimates robust DID bounds on the ATTs for each of post-treatment periods under the canonical $2 \times 2$ DID setting.
\vspace{1mm}
\end{minipage}\\

\begin{minipage}{.2\textwidth}
\texttt{rdidstag}
\end{minipage}
&
\begin{minipage}{.7\textwidth}
\vspace{1mm}
Estimates robust bounds on the cohort-time ATTs under the staggered adoption design.
\vspace{1mm}
\end{minipage}\\

\hline  
\end{tabular}
\renewcommand{\arraystretch}{1}
\end{table}

Our article contributes to the expanding suite of publicly available software within the DID design community, including \cite{villa2016diff}, \cite{houngbedji2016abadie}, \cite{de2019fuzzy}, \cite{rios2022drdid, rios2023csdid}, among others. 
Specifically, our software is more closely aligned with research exploring alternative identifying assumptions within the DID framework; for instance, \cite{mora2015didq} examines DID estimations under alternative assumptions and assesses their robustness, while \cite{bravo2022honestdid} discusses methods to bound post-treatment parallel trends (PT) violations using pre-treatment PT violations.

The rest of this article is structured as follows.
In section 2, we recap the underlying frameworks of the robust DID bounds by \cite{bk2023rdid}.
In sections 3, 4, and 5, we describe the \texttt{rdid}, \texttt{rdid\_dy}, \texttt{rdidstag} commands, respectively.
In section 6, we illustrate the performance of each command through a series of Monte Carlo simulations.

\section{Framework and methodology}

Consider the following potential outcome model:
\begin{eqnarray}\label{seq1}
Y_t&=&Y_t(1)D+Y_t(0)(1-D)
\end{eqnarray}
where $(Y_t,D)$, $t\in \mathcal T_0 \cup \{1\}$ represents the observed data, while the vector $(Y_1(0), Y_1(1))$ is latent. 
The variables $Y_{t_0}, Y_1 \in \mathcal Y$ are respectively the observed outcomes in the baseline period $t_0 \in \mathcal T_0$ and the follow-up period 1, while $D\in \left\{0,1\right\}$ is the observed treatment that occurred between periods 0 and 1, $Y_1(0)$ and $Y_1(1)$ are the potential outcomes that would have been observed in period 1 had the treatment $D$ been externally set to 0 and 1, respectively. $\mathcal T_0$ denotes the set of all pre-treatment periods. 
Model (\ref{seq1}) assumes that there is no anticipatory effect of the treatment, so that $Y_{t_0}(1)=Y_{t_0}(0)$ for all baseline periods $t_0 \in \mathcal T_0$.

The standard DID estimand is defined as the difference between the OLS estimand in period 1 and the selection bias in period 0: $$\theta_{DID} \equiv \theta_{OLS}-SB_0$$ where $\theta_{OLS}$ is the coefficient of a regression of $Y_1$ on the binary $D$,
$$
\theta_{OLS} = \mathbb{E}[Y_1 \vert D=1 ] - \mathbb{E}[Y_1 \vert D=0 ],
$$
and $SB_t$ is the selection bias in period $t$,
\begin{align*}
SB_t &= \mathbb{E}[Y_t(0) \vert D=1] - \mathbb{E}[Y_t(0) \vert D=0],
\end{align*}
for $t\in \mathcal T_0 \cup \{1\}$.
Recall that $SB_0$ is identified with the no anticipatory effect but $SB_1$ is an unobserved counterfactual.

Note that the average treatment effect on the treated (ATT),
$$ATT\equiv \mathbb E[Y_1(1)-Y_1(0)\vert D=1],$$
is identified by $\theta_{DID}$ under the parallel trend (PT) assumption,
$$
\mathbb{E}[Y_1(0) - Y_0(0) \vert D=1] = \mathbb{E}[Y_1(0) - Y_0(0) \vert D=0],
$$
or $SB_1 = SB_0$ (bias stability).

Hence, the identification of ATT can be re-framed as imputing the counterfactual selection bias $SB_1$. This observation inspires us to introduce a robust version of the DID estimand. 

Define the baseline information set $\mathcal I_0\equiv \left\{(t_0,x_{t_0}): x_{t_0} \in \mathcal X_{t_0}, t_0 \in \mathcal T_0\right\}$, where $\mathcal X_{t_0}$ is the support of a baseline covariate $X_{t_0}$ in period $t_0$, and $\mathcal T_0$ is the set of pre-treatment periods. For simplicity, suppose there are no baseline covariates so that $\mathcal I_0=\mathcal T_0$.

\begin{definition}
Given the baseline information set $\mathcal I_0=\mathcal T_0$, we define the \textit{robust difference-in-differences (RDID)} estimand as
\begin{eqnarray}\label{RDID}
\theta_{RDID}\equiv \theta_{OLS}-\operatorname{Conv}\big(SB(\mathcal T_0)\big),
\end{eqnarray}
where $\operatorname{Conv}(A)$ is the convex hull of a generic set $A$, $SB$ is a function defined as
\begin{eqnarray*}
SB\colon \quad \mathcal I_0& \to & \mathbbm R,\\
                   \iota_0& \mapsto &  SB(t_0)=SB_{t_0},
\end{eqnarray*}
and $SB(\mathcal T_0)$ is the image of the information set $\mathcal I_0=\mathcal T_0$ through the function $SB$. 
\end{definition}
When baseline covariates exists, we define the function $SB$ as 
$$
\begin{array}{llcl}
SB\colon 	&   \mathcal I_0 &\to& \mathbbm R,\\
			& \iota_0 \equiv (t_0, x_{t_0}) &\mapsto& SB(\iota_0)\equiv \mathbb E[Y_{t_0} \vert D=1, X_{t_0}=x_{t_0}]-\mathbb E[Y_{t_0} \vert D=0, X_{t_0}=x_{t_0}].
\end{array}
$$

In the above definition, if there is only one element in the information, i.e., $\mathcal I_{0}=\mathcal T_0=\{0\}$, then $\operatorname{Conv}\big(SB(\mathcal T_0)\big)=SB_0$, and the robust DID estimand is the same as the standard DID estimand. 

\subsection{Bias set stability}

Let us first consider the simple case with no covariates in the model. %
\begin{assumption}[Bias set stability]\label{sb:bounds}
\begin{eqnarray*}
SB_1 \in \left[ \inf_{\iota_0\in \mathcal I_0} SB(\iota_0), \sup_{\iota_0 \in \mathcal I_0} SB(\iota_0)\right] \equiv \Delta_{SB(\mathcal I_0)},
\end{eqnarray*}
where $SB(\iota_0)$ is defined above. 
\end{assumption} 

Assumption \ref{sb:bounds} is weaker than the standard ``parallel/common trends'' assumption. Indeed, if $\mathcal{I}_0$ is the singleton of a single baseline information $I_0=\{0\}$, then Assumption~\ref{sb:bounds} is equivalent to 
$SB_1= SB_0,$ which is equivalent to the parallel trends assumption.

Under Assumption \ref{sb:bounds}, robust DID estimand yields the following bounds on the ATT.\footnote{Note that the RDID can be generalized by using other correspondences $f(SB(\mathcal I_0))$ than $\operatorname{Conv}(SB(\mathcal I_0))$ associated with other assumptions \citep{bk2023rdid}.}
\begin{proposition}\label{prop1}
Suppose that model (\ref{seq1}) along with Assumption \ref{sb:bounds} holds. Then, the following bounds hold for the ATT: 
\begin{eqnarray*}
ATT \in \left[ \theta_{OLS}- \sup_{\iota_0 \in \mathcal I_0} SB(\iota_0), \theta_{OLS}- \inf_{\iota_0 \in \mathcal I_0} SB(\iota_0) \right] \equiv \Theta_I.
\end{eqnarray*} 
These bounds are sharp, and $\Theta_I$ is the identified set for the ATT.
\end{proposition}

The bounds in Proposition \ref{prop1} are never empty, as they always contain the standard DID estimand under the parallel trends assumption. However, they may not contain the OLS estimand in period 1, $\theta_{OLS}$, as 0 may not lie within the set $\Delta_{SB(\mathcal I_0)}$. If all pre-treatment periods selection biases are equal, i.e., $SB(\iota_0)=SB_0$ for all $\iota_0$, then our bounds collapse to a point, the standard DID estimand. In case the information set $\mathcal I_0$ is the set of pre-treatment periods, the above bounds are robust to violations of PT that can be captured in the pre-treatment periods.

\subsection{PO-RDID: Best predictor of $SB_1$ based on a loss function}
Consider a random variable $I_0$ representing the baseline information with a distribution $\mu_{I_0}$. %
Provided $\mu_{I_0}$ is known, we are going to assume that the decision maker will choose the selection bias $SB_1$ in such a way that a loss function is minimized. By plugging such an optimal selection bias $SB_1$ into the definition of the robust DID, we obtain what we call a \textit{policy-oriented robust difference-in-differences (PO-RDID)} estimand. This estimand may not have a causal interpretation, but it may help the policy-maker in her decision making process. 

\begin{assumption}\label{ass:optimalDID}
Let $\mathcal L(b; SB(\mathcal I_0), \mu_{I_0})$ be the decision maker's loss function when she observes a (random) baseline information $I_0$ and chooses a selection bias $b$.
The decision maker chooses $b$ (or $SB_1$) to minimize the loss $\mathcal L(b)$: $SB_1^{opt}=\arg\min_b \mathcal L(b; SB(\mathcal I_0), \mu_{I_0})$.
\end{assumption}
In this paper, we consider the class of $p$-norm losses defined as: $$\mathcal L_p(b; SB(\mathcal I_0), \mu_{I_0})=\left(\mathbb E_{\mu_{I_0}}\left[\vert b - SB(I_0)\vert ^p \right]\right)^{1/p},$$ where $1 \leq p \leq \infty$.
We are going to derive the optimal selection bias $SB_1^{opt}$ for $p\in \left\{1,2,\infty\right\}$. We consider those special loss functions because the solutions to the optimization problem have closed-form expressions. Other loss functions can also be considered. 

The command \texttt{rdid} with an option \texttt{rdidtype(1)} can be used, and the distribution of $SB_0(I_0)$ is estimated by the proportion of observations at each level of $I_0$.

\subsubsection{$L1$ loss: Mean absolute error (MAE)} 
Note that $\mathcal L_1(b; SB(\mathcal I_0), \mu_{I_0})=\mathbb E_{\mu_{I_0}}\left[\vert b - SB(I_0)\vert \right]$.
Given this $L1$ loss function, under Assumption \ref{ass:optimalDID}, the decision maker solves the following optimization problem:
\begin{eqnarray*}
\min_{SB_1} \mathbb E_{\mu_{I_0}}\left[\vert SB_1 - SB(I_0)\vert \right].
\end{eqnarray*}
The optimal decision is to set the selection $SB_1$ to be equal to the median selection bias in the baseline period, i.e., $SB_1^{opt}=Med_{\mu_{I_0}}\big(SB(I_0)\big)$. 
In such a case, the policy-oriented robust DID estimand is given by $$\theta_{PO-RDID}^{\mathcal{L}_{1}}=\theta_{OLS}-Med_{\mu_{I_0}}\big(SB(I_0)\big).$$

\subsubsection{$L2$ loss: Root mean square error (RMSE)}
We have $\mathcal L_2(b; SB(\mathcal I_0), \mu_{I_0})=\left(\mathbb E_{\mu_{I_0}}\left[\vert b - SB_0(I_0)\vert ^2 \right]\right)^{1/2}$, and minimizing the RMSE is equivalent to minimizing the mean square error (MSE). Therefore, under Assumption~\ref{ass:optimalDID}, the decision maker solves the following optimization problem:
\begin{eqnarray*}
\min_{SB_1} \mathbb E_{\mu_{I_0}}\left[\left(SB_1 - SB(I_0)\right)^2 \right].
\end{eqnarray*}
This yields an optimal decision for the selection $SB_1$ to be set equal to the average selection bias in the baseline period, i.e., $SB_1^{opt}=\mathbb E_{\mu_{I_0}}[SB(I_0)]$. Hence, we have $$\theta_{PO-RDID}^{\mathcal{L}_{2}}=\theta_{OLS}-\mathbb E_{\mu_{I_0}}[SB(I_0)].$$

\subsubsection{$L\infty$ loss: Maximal regret} 
Note that
\begin{align*}
\mathcal L_{\infty}(b; SB(\mathcal I_0), \mu_{I_0}) &= \text{ess}\sup \vert b - SB(I_0)\vert, \\
&= \inf \Big\{ M \geq 0: \mathbbm{P}_{\mu_{I_0}} \big( \vert b - SB(I_0)\vert \leq M \big) = 1 \Big\}.
\end{align*}

This optimization problem with the $L\infty$ loss is equivalent to a minimax criterion, and yields the mid-point of the bounds on $SB_1$ stated in Assumption \ref{sb:bounds}. Hence, the PO-RDID estimand is given by
$$\theta_{PO-RDID}^{\mathcal{L}_{\infty}}=\theta_{OLS}-\frac{1}{2}\bigg(\inf_{\iota_0 \in \mathcal I_0}SB(\iota_0)+\sup_{\iota_0 \in \mathcal I_0}SB(\iota_0)\bigg).$$

Note that in all cases, if the information in the baseline period is a singleton, then the optimal $SB_1^{opt}$ is the selection bias in the baseline period $SB_0$, which is equivalent to the parallel trends assumption. 
Unlike the PO-RDID estimand obtained from $L1$ and $L2$ loss functions, that obtained from the $L\infty$ loss function does not require the knowledge of the distribution $\mu_{I_0}$ of the information $I_0$ but only its support and is easy to compute. 
However, when the distribution of $SB(I_0)$ is uniform over $\left[ \inf_{\iota_0\in \mathcal I_0} SB(\iota_0), \sup_{\iota_0 \in \mathcal I_0} SB(\iota_0)\right]$, then the optimal selection bias $SB_1^{opt}$ is the same in all three cases.

\subsection{Forecasting $SB_1$ when the baseline information is ordered}\label{sec:orderedinfo}
Suppose that the baseline information set $\mathcal I_0$ is ordered (e.g., a set of multiple pre-treatment periods $\mathcal I_0=\{-T_0, -T_0+1, \ldots, -1, 0\}$ or a continuous baseline covariate $X_0$ like \textit{age}). We can regress $SB(I_0)$ on $\{I_0, I_0^2, \ldots\}$ and use this regression to predict $SB_1$.

For instance, if $\mathcal I_0=\{-T_0, -T_0+1, \ldots, -1, 0\}$ and the selection bias $SB(I_0)$ is increasing over time, Assumption \ref{sb:bounds} may not hold. The researcher could instead use this increasing trend information about the selection bias to forcast the next period selection bias $\widehat{SB}_1$.

The command \texttt{rdid} with an option \texttt{rdidtype(2)} can be used for a linear predicition model, and an option \texttt{peval(\#)} can be used to specify the evaluation point of the prediction.

\subsection{Identification with multiple periods}

We generalize our analysis to a setting where the treatment receipt occurs at multiple periods. We consider the following multiple treatment periods potential outcome model:
\begin{eqnarray}\label{eq:ext}
Y_{t} &=&\sum_{(d_1,\ldots, d_T) \in \{0,1\}^T} Y_{t}(0,d_1, \ldots, d_T) \mathbbm{1}\{D_0=0,D_1=d_1, \ldots, D_T=d_T\},%
     \end{eqnarray} 
for $t\in \mathcal T_0 \cup \{1,\ldots,T\}$, where $Y_t$ denotes the observed outcome in period $t$, $D_t$ is the observed treatment status in period $t$ with $D_0=0$ by definition, while $Y_t(0,d_1, \ldots, d_T)$ is the potential outcome when the treatment path $(D_0,D_1, \ldots, D_T)$ is externally set to $(0,d_1,\ldots, d_T)$.\footnote{See \cite{Robins1986, Robins1987}, and \cite{Han2021} for a similar definition of the potential outcome model.} 
Under the no-anticipation assumption, we have $Y_{t_0}(0,d_1,\ldots, d_T) =Y_{t_0}(0)$ for all $(d_1,\ldots,d_T) \in \{0,1\}^T$ and $t_0 \in \mathcal T_0$.
That is, we assume that individuals do not anticipate any effects of the treatment before it occurs for the first time in $t=1$.
However, we allow the individuals to anticipate the effects of the treatment for the rest of the period. 
This assumption is less restrictive than the commonly used no-anticipatory effects assumption.

In the above framework, the parameter of interest is the average treatment effect in period $t$ on the treated group following the path $\bm d^0 \equiv (0,d_1^0,\ldots, d_T^0)$ to $\bm d^1 \equiv (0,d_1^1,\ldots, d_T^1)$, which is defined as:%
\begin{eqnarray*}
&& ATT_t[\bm d^0 \rightarrow \bm d^1]\\
&& \qquad \qquad \equiv \mathbb E\left[Y_t(\bm d^1)-Y_t(\bm d^0) \vert (D_0,D_1, \ldots, D_T)=\bm d^1\right].
\end{eqnarray*}

Similarly to what we have in the one post-treatment setting, we can write the difference-in-means estimand $(\theta_{DIM}^t)$ between the two groups $\bm d^0$ and $\bm d^1$ in period $t$ as: 
\begin{align*}
\theta_{DIM}^t[\bm d^0 \rightarrow \bm d^1]&\equiv \mathbb E\left[Y_t \vert (D_0,D_1, \ldots, D_T)=\bm d^1\right]-\mathbb E\left[Y_t \vert (D_0,D_1, \ldots, D_T)=\bm d^0\right]\\
& =ATT_t[\bm d^0\rightarrow \bm d^1]+SB_t[\bm d^0\rightarrow \bm d^1], 
\end{align*}
where $SB_t[\bm d^0\rightarrow \bm d^1]\equiv \mathbb E\left[Y_t(\bm d^0) \vert (D_0,D_1, \ldots, D_T)=\bm d^1\right]-\mathbb E\left[Y_t(\bm d^0) \vert (D_0,D_1, \ldots, D_T)=\bm d^0\right]$.
We extend Assumption \ref{sb:bounds} to the current setting. 

\begin{assumption}[Extended bias set stability]\label{sb:boundsex}
For each $t$, 
\begin{align*}
SB_t[\bm d^0\rightarrow \bm d^1] &\in \left[ \inf_{t_0\in \mathcal T_0} SB_{t_0}[\bm d^0\rightarrow \bm d^1], \sup_{t_0 \in \mathcal T_0} SB_{t_0}[\bm d^0\rightarrow \bm d^1]\right] \\
&\equiv \Delta_{SB(\mathcal T_0)}[\bm d^0\rightarrow \bm d^1],
\end{align*}
where $SB_{t_0}[\bm d^0\rightarrow \bm d^1]\equiv \mathbb E[Y_{t_0}(0)\vert (D_0,D_1, \ldots, D_T)=\bm d^1] - \mathbb E[Y_{t_0}(0)\vert (D_0,D_1, \ldots, D_T)=\bm d^0]$ is the baseline selection bias with respect to the treatment status in period $t_0 \in \mathcal T_0$ when the information set $\mathcal I_0$ is equal to $\mathcal T_0$. 
\end{assumption} 
Assumption \ref{sb:boundsex} is a generalization of Assumption \ref{sb:bounds}, and we have the following identification results for $ATT_t$ with different treatment paths.

\begin{proposition}\label{prop1ex}
Suppose that model (\ref{eq:ext}) along with Assumption \ref{sb:boundsex} holds. Then, the following bounds are valid for $ATT_t$: 
\begin{align*}
&ATT_t[\bm d^0 \rightarrow \bm d^1] \\
&\qquad \in \left[ \theta_{DIM}^t[\bm d^0\rightarrow \bm d^1] - \sup_{t_0 \in \mathcal T_0} SB_{t_0}[\bm d^0\rightarrow \bm d^1], \theta_{DIM}^t[\bm d^0\rightarrow \bm d^1]- \inf_{t_0 \in \mathcal T_0} SB_{t_0}[\bm d^0\rightarrow \bm d^1] \right], \\
& \qquad \equiv \Theta_I^t[\bm d^0 \rightarrow \bm d^1].
\end{align*}
These bounds are sharp, and $\Theta_I^t[\bm d^0 \rightarrow \bm d^1]$ is the identified set for $ATT_t[\bm d^0 \rightarrow \bm d^1]$.
\end{proposition}

In a staggered design setting, the average treatment effect in period $t$ on units who are treated for the first time in period $g$ could be an interesting parameter, as considered in \cite{CallawaySantAnna2021}:
 $$ATT(g, t) \equiv ATT_t[(0,\ldots,0,d_g=0,0,\ldots, 0)\rightarrow (0,\ldots,0,d_g=1,1,\ldots, 1)].$$  

By defining $G$ as
$$
G = \left\{\begin{array}{ccl}
g & \mathrm{if} & (D_{0}, D_{1}, \ldots, D_{T}) = (0, \ldots, 0, d_g = 1, 1, \ldots, 1) \\
\infty & \mathrm{if} & (D_{0}, D_{1}, \ldots, D_{T}) = (0, \ldots, 0)
\end{array}\right. ,
$$
where $G = \infty$ denotes the never-treated cohort as our control group, we can write
\begin{align*}
\theta^t_{DIM}[\infty \rightarrow g] &\equiv  \theta^t_{DIM}[(0, \ldots, 0)\rightarrow (0, \ldots, 0, d_g = 1, 1, \ldots, 1)] \\
&= \mathbb{E}[Y_t \vert G = g] - \mathbb{E}[Y_t \vert G = \infty] \\
SB_t[\infty \rightarrow g] &\equiv SB_t[(0, \ldots, 0)\rightarrow (0, \ldots, 0, d_g = 1, 1, \ldots, 1)] \\
	&= \mathbb{E}[Y_t(g) \vert G = g] - \mathbb{E}[Y_t(\infty) \vert G = \infty],
\end{align*}
for $t \in \mathcal T_0 \cup \{1, \ldots, T \}$, by abusing the notations of $g$ and $\infty$ for paths.

Hence, Assumption \ref{sb:boundsex} can be re-written as
\begin{align*}
SB_t[\infty \rightarrow g] &\in \left[ \inf_{t_0\in \mathcal T_0} SB_{t_0}[\infty \rightarrow g], \sup_{t_0 \in \mathcal T_0} SB_{t_0}[\infty \rightarrow g]\right] \\
&\equiv \Delta_{SB(\mathcal T_0)}[\infty \rightarrow g],
\end{align*}
for each $t \in \{1, \ldots, T \}$, and $SB_{t_0}[\infty \rightarrow g] \equiv \mathbb E[Y_{t_0}(0)\vert G = g] - \mathbb E[Y_{t_0}(0)\vert G = \infty ]$ for $t \in \mathcal T_0$.
Thus, we have 
\begin{align*}
ATT(g, t) &\in \left[ \theta_{DIM}^t[\infty \rightarrow g] - \sup_{t_0 \in \mathcal T_0} SB_{t_0}[\infty \rightarrow g], \theta_{DIM}^t[\infty \rightarrow g]- \inf_{t_0 \in \mathcal T_0} SB_{t_0}[\infty \rightarrow g] \right].\\
&\equiv \Theta(g, t)
\end{align*}

\section{The rdid command}
\subsection{Syntax}

The \texttt{rdid} command estimates bounds on the ATT under the assumptions discussed above.
The syntax for the \texttt{rdid} command is

\begin{stsyntax}
rdid
    \depvar\
    \optindepvars\
    \optif\
    \optin ,
    \underbar{treat}name(\varname)
    \underbar{post}name(\varname)
    \underbar{info}name(\varname)
    \optional{
    rdidtype(\num)
    peval(\num)
    \underbar{lev}el(\num)
    \underbar{fig}ure
    \underbar{b}rep(\num)
    \underbar{cl}ustername(\varname)}
\end{stsyntax}

\subsection{Options}

\hangpara
{\tt treatname(\varname)} specifies the variable name of the treatment indicator. \texttt{treatname()} is required.

\hangpara
{\tt postname(\varname)} specifies the variable name of the post-treatment period indicator. \texttt{postname()} is required.

\hangpara
{\tt infoname(\varname)} specifies the variable name of the information index. \texttt{infoname()} is required.

\hangpara
{\tt rdidtype(\num)} specifies the type of RDID estimator; \texttt{0} for the simple RDID, \texttt{1} for the policy-oriented (PO) RDID, and \texttt{2} for the RDID with linear predictions. The default is {\tt rdidtype(0)}.

\hangpara
{\tt peval(\num)} specifies the evaluation point for the RDID with linear predictions. The default is the mean value of {\tt infoname(\varname)} in the post-treatment period. This option is ignored when {\tt rdidtype(\num)} is not equal to \texttt{2}.

\hangpara
{\tt level(\num)} specifies the confidence level, as a percentage,
for confidence intervals (CIs). The default is {\tt level(95)}.
Three different CIs are provided with {\tt rdidtype(0)}; 1) CI for the bounds \citep{ye2023negative}, 2) CI for the ATT \citep{ye2023negative}, and 3) CI for the bounds (union bounds).
Bootstrapped CIs are provided with {\tt rdidtype(1)} and {\tt rdidtype(2)}.

\hangpara
When {\tt figure(\ststring)} is specified, \texttt{rdid} saves a scatter plot of estimated selection biases and the information index in the pre-treatment periods as {\tt figure(\ststring).png}.

\hangpara
{\tt brep(\num)} specifies the number of bootstrap replicates. The default is {\tt brep(500)}.

\hangpara
{\tt clustername(\varname)} specifies the variable name that identifies resampling clusters in bootstrapping.

\subsection{Stored results}

\texttt{rdid} stores the following in \texttt{e()}. 
When \textit{indepvars} is specified, $\tau^{DR}$ is estimated using post-treatment periods (see \citeauthor{bk2023rdid}, \citeyear{bk2023rdid}) and stored in {\tt e(DR)}. Otherwise, $\theta_{OLS}$ is estimated and stored in {\tt e(OLS)}.
Also, stored scalars depend on {\tt rdidtype(\num)} specification (referred as `type' below). 
In the following, LB (UB) stands for lower bound (upper bound).

\begin{stresults}
\stresultsgroup{Scalars (common)} \\
\stcmd{e(DR)} & $\widehat{\tau}^{DR}$ estimate
&
\stcmd{e(OLS)} & $\widehat{\theta}_{OLS}$ estimate
\\
\stcmd{e(N)} & number of observations
&
&
\\
\stresultsgroup{Scalars (type=0)} \\
\stcmd{e(SB\_LB)} & LB of $\widehat{\Delta}_{SB(\mathcal I_0)}$
&
\stcmd{e(SB\_UB)} & UB of $\widehat{\Delta}_{SB(\mathcal I_0)}$
\\
\stcmd{e(RDID\_LB)} & LB of $\widehat{\Theta}_I$
&
\stcmd{e(RDID\_UB)} & UB of $\widehat{\Theta}_I$
\\
\stcmd{e(CI1\_LB)} & LB of CI-1
&
\stcmd{e(CI1\_UB)} & UB of CI-1
\\
\stcmd{e(CI2\_LB)} & LB of CI-2
&
\stcmd{e(CI2\_UB)} & UB of CI-2
\\
\stcmd{e(CI3\_LB)} & LB of CI-3
&
\stcmd{e(CI3\_UB)} & UB of CI-3
\\
\stresultsgroup{Scalars (type=1)} \\
\stcmd{e(L1\_CI\_LB)} & LB of CI for $\widehat{\theta}_{PO-RDID}^{\mathcal{L}_1}$
&
\stcmd{e(L1\_CI\_LB)} & UB of CI for $\widehat{\theta}_{PO-RDID}^{\mathcal{L}_1}$
\\
\stcmd{e(L2\_CI\_LB)} & LB of CI for $\widehat{\theta}_{PO-RDID}^{\mathcal{L}_2}$
&
\stcmd{e(L2\_CI\_LB)} & UB of CI for $\widehat{\theta}_{PO-RDID}^{\mathcal{L}_2}$
\\
\stcmd{e(Linf\_CI\_LB)} & LB of CI for $\widehat{\theta}_{PO-RDID}^{\mathcal{L}_{\infty}}$
&
\stcmd{e(Linf\_CI\_LB)} & UB of CI for $\widehat{\theta}_{PO-RDID}^{\mathcal{L}_{\infty}}$
\\
\stcmd{e(L1\_PE)} & $\widehat{\theta}_{PO-RDID}^{\mathcal{L}_1}$ estimate
&
\stcmd{e(L2\_PE)} & $\widehat{\theta}_{PO-RDID}^{\mathcal{L}_2}$ estimate
\\
\stcmd{e(Linf\_PE)} & $\widehat{\theta}_{PO-RDID}^{\mathcal{L}_{\infty}}$ estimate
&
&
\\
\stresultsgroup{Scalars (type=2)} \\
\stcmd{e(SB\_hat)} & $\widehat{SB}_1$ estimate
&
\stcmd{e(proj\_PE)} & $\widehat{\theta}_{RDID}$ estimate
\\
\stcmd{e(CI\_LB)} & LB of CI for $\widehat{\theta}_{RDID}$
&
\stcmd{e(CI\_UB)} & UB of CI for $\widehat{\theta}_{RDID}$
\\
\stresultsgroup{Macros} \\
\stcmd{e(cmd)} & \stcmd{rdid}
&
\stcmd{e(indep)} & name of independent variable
\\
\stcmd{e(depvar)} & name of dependent variable(s)
&
\stcmd{e(clustvar)} & name of cluster variable
\\
\stcmd{e(level)} & confidence level
&
&
\\
\stresultsgroup{Matrices} \\
\stcmd{e(results)} & displayed summary matrix
&
&
\\
\end{stresults}

\section{The rdid\_dy command}

\subsection{Syntax}

The \texttt{rdid\_dy} command is a wrapper command that implements RDID estimation separately for each levels of a time variable in the post-treatment period and collects the results.
The syntax for the \texttt{rdid\_dy} command is

\begin{stsyntax}
rdid\_dy
    \depvar\
    \optindepvars\
    \optif\
    \optin ,
    \underbar{treat}name(\varname)
    \underbar{post}name(\varname)
    \underbar{info}name(\varname)
    \underbar{t}name(\varname)
    \optional{
    rdidtype(\num)
    peval(\num)
    \underbar{lev}el(\num)
    \underbar{fig}ure(\ststring)
    \underbar{b}rep(\num)
    \underbar{cl}ustername(\varname)
    \underbar{ci}type(\num)
    \underbar{l}osstype(\num)}
\end{stsyntax}

\subsection{Options}

\hangpara
{\tt tname(\varname)} specifies the variable name for the time. \texttt{tname()} is required.

\hangpara
{\tt peval(\num)} specifies the evaluation point for the RDID with linear predictions. 
The default is the mean value of {\tt infoname(\varname)} in the post-treatment period. 
In particular, if {\tt infoname(\varname)} is the same as {\tt tname(\varname)}, each RDID estimate is obtained using each level of {\tt tname(\varname)} as the evaluation point. 
This option is ignored when {\tt rdidtype(\num)} is not equal to \texttt{2}.

\hangpara
When {\tt figure(\ststring)} is specified, \texttt{rdid\_dy} saves line plots of RDID estimates and confidence intervals over the post-treatment periods as {\tt figure(\ststring).png}.

\hangpara
{\tt citype(\varname)} specifies the type of confidence intervals to collect for {\tt rdidtype(0)}; \texttt{1} yields CI for the RDID bounds \citep{ye2023negative}, \texttt{2} yields CI for the ATT \citep{ye2023negative}, and \texttt{3} yields CI for the RDID bounds (Union Bounds). The default is {\tt citype(1)}. This option is ignored when {\tt rdidtype(\num)} is not equal to \texttt{0}. 

\hangpara
{\tt losstype(\num)} specifies the type of loss function for {\tt rdidtype(1)} (PO-RDID); \texttt{1} for $\mathcal{L}_1$, \texttt{2} for $\mathcal{L}_2$, and \texttt{0} for $\mathcal{L}_{\infty}$. 
The default is {\tt losstype(1)}. This option is ignored when {\tt rdidtype(\num)} is not equal to \texttt{1}.

\subsection{Stored results}

\texttt{rdid\_dy} stores the following in \texttt{e()}. 
Stored scalars depend on {\tt rdidtype(\num)} specification; 
In the following, LB (UB) stands for lower bound (upper bound) and (t) denotes each level of {\tt tname(\varname)}.

\begin{stresults}
\stresultsgroup{Scalars} \\
\stcmd{e(N)} & \qquad number of observations
&
\stcmd{e(RDID\_PE\_(t))} & \qquad $\widehat{\theta}_{RDID}$ estimate for (t)
\\
\stcmd{e(RDID\_LB\_(t))} & \qquad LB of $\widehat{\Theta}_{I}$ for (t)
&
\stcmd{e(RDID\_UB\_(t))} & \qquad UB of $\widehat{\Theta}_{I}$ for (t)
\\
\stcmd{e(RDID\_LB\_(t))} & \qquad LB of CI for (t)
&
\stcmd{e(RDID\_UB\_(t))} & \qquad LB of CI for (t)
\\
\stresultsgroup{Macros} \\
\stcmd{e(cmd)} & \qquad \stcmd{rdid\_dy}
&
\stcmd{e(indep)} & \qquad name of independent
\\
&&& \qquad variable
\\
\stcmd{e(depvar)} & \qquad name of dependent 
&
\stcmd{e(clustvar)} & \qquad name of cluster variable
\\
& \qquad variable(s)&&
\\
\stcmd{e(level)} & \qquad confidence level
&
&
\\
\stresultsgroup{Matrices} \\
\stcmd{e(results)} & \qquad displayed summary matrix
&
&
\\
\end{stresults}

\section{The rdidstag command}
\subsection{Syntax}

The \texttt{rdidstag} command estimates bounds on the ATT over time for different cohorts in the staggered adoption design under the assumptions discussed above.
The syntax for the \texttt{rdidstag} command is

\begin{stsyntax}
rdidstag
    \depvar\
    \optindepvars\
    \optif\
    \optin ,
    \underbar{g}name(\varname)
    \underbar{t}name(\varname)
    \optional{
    \underbar{post}name(\varname)
    \underbar{info}name(\varname)
    \underbar{lev}el(\num)
    \underbar{fig}ure
    \underbar{cl}ustername(\varname)}
\end{stsyntax}

\subsection{Options}

\hangpara
{\tt gname(\varname)} specifies the variable name of the cohort index, $g \in \mathcal{G}$, where $g=0$ is assigned for the never-treated group. \texttt{gname()} is required.

\hangpara
{\tt tname(\varname)} specifies the variable name of time index, $t$. \texttt{tname()} is required.

\hangpara
{\tt postname(\varname)} specifies the variable name of the post-treatment period indicator. The default is using units with $t \geq \min(\mathcal{G} \setminus \{0\})$ as post-period units. 

\hangpara
{\tt infoname(\varname)} specifies the variable name of the information index. The default is using pre-treatment periods; i.e., $\mathcal{I}_0 = \{t \mid t < \min(\mathcal{G} \setminus \{0\}) \}$. 

\hangpara
{\tt level(\num)} specifies the confidence level, as a percentage,
for confidence intervals. The default is {\tt level(95)}.

\hangpara
When {\tt figure} is specified, \texttt{rdidstag} saves line plots of RDID bounds and confidence intervals over the post-treatment periods for each $g \in \mathcal{G}$ as {\tt figure(\ststring).png}.

\hangpara
{\tt clustername(\varname)} specifies the variable name that identifies resampling clusters in bootstrapping.

\subsection{Stored results}

\texttt{rdidstag} stores the following in \texttt{e()}.
In the following, LB (UB) stands for lower bound (upper bound).
Also, (g) and (t) denote each level of {\tt gname(\varname)} and {\tt tname(\varname)}, respectively.

\begin{stresults}
\stresultsgroup{Scalars} \\
\stcmd{e(RDID\_LB\_(g)\_(t))} & \qquad LB of $\widehat{\Theta}(g, t)$
&
\stcmd{e(RDID\_UB\_(g)\_(t))} & \qquad UB of $\widehat{\Theta}(g, t)$
\\
\stcmd{e(CI\_LB\_(g)\_(t))} & \qquad LB of CI for $\widehat{\Theta}(g, t)$
&
\stcmd{e(CI\_UB\_(g)\_(t))} & \qquad UB of CI for $\widehat{\Theta}(g, t)$
\\
\stcmd{e(N)} & \qquad number of observations
&
&
\\
\stresultsgroup{Macros} \\
\stcmd{e(cmd)} & \qquad \stcmd{rdidstag}
&
\stcmd{e(indep)} & \qquad name of independent
\\
&&& \qquad variable
\\
\stcmd{e(depvar)} & \qquad name of dependent 
&
\stcmd{e(clustvar)} & \qquad name of cluster variable
\\
& \qquad variable(s)&&
\\
\stcmd{e(level)} & \qquad confidence level
&
&
\\
\stresultsgroup{Matrices} \\
\stcmd{e(results)} & \qquad displayed summary matrix
&
&
\\
\end{stresults}

\section{Applications}
We illustrate usages of \texttt{rdid}, \texttt{rdid\_dy}, and \texttt{rdidstag} commands using two applications.
We first demonstrate \texttt{rdid} and \texttt{rdid\_dy} commands using an application of \cite{cai2016} in Section \ref{sec.cai} and then \texttt{rdidstag} command using data from \cite{dube2016} in Section \ref{sec.dube}.

\subsection{Effects of insurance provision on tobacco production}\label{sec.cai}

\cite{cai2016} investigates the impact of insurance provision on tobacco production using a household-level panel dataset provided by the Rural Credit Cooperative (RCC), the main rural bank in China.

\subsubsection{Using rdid command}

In the following runs, the dependent variable is \texttt{area\_tob} and the covariates are \texttt{hhsize}, \texttt{educ\_scale}, and \texttt{age}, yielding the doubly-robust estimate \citep{bk2023rdid}.
Also, treatment variable and the post-treatment period indicator is is \texttt{treatment} and \texttt{policy2}, respectively, and the information set is specified as pre-treatment periods by using \texttt{year}.
We use 1,000 bootstrap replicates and cluster the standard errors using \texttt{hhno}.
First, the following is yielding the RDID bounds with the three different confidence intervals.

\begin{stlog}
\input{logs/log_rdid_0.log.tex}\nullskip
\end{stlog}

Second, by specifying the option {\tt rdidtype(1)}, we can have PO-RDID estimates and their confidence intervals under the three loss functions.

\begin{stlog}
\input{logs/log_rdid_1.log.tex}\nullskip
\end{stlog}

Lastly, by specifying the option {\tt rdidtype(2)}, we can have an RDID estimate based on a linear prediction with an ordered information set, pre-treatment periods.
The predicted value is evaluated at the mean of the post-treatment periods, \texttt{2005.5}, with the option {\tt peval(2005.5)}.

\begin{stlog}
\input{logs/log_rdid_2.log.tex}\nullskip
\end{stlog}

\subsubsection{Using rdid\_dy command}

\begin{figure}[h]
	\centering
	\includegraphics[width=0.8\textwidth]{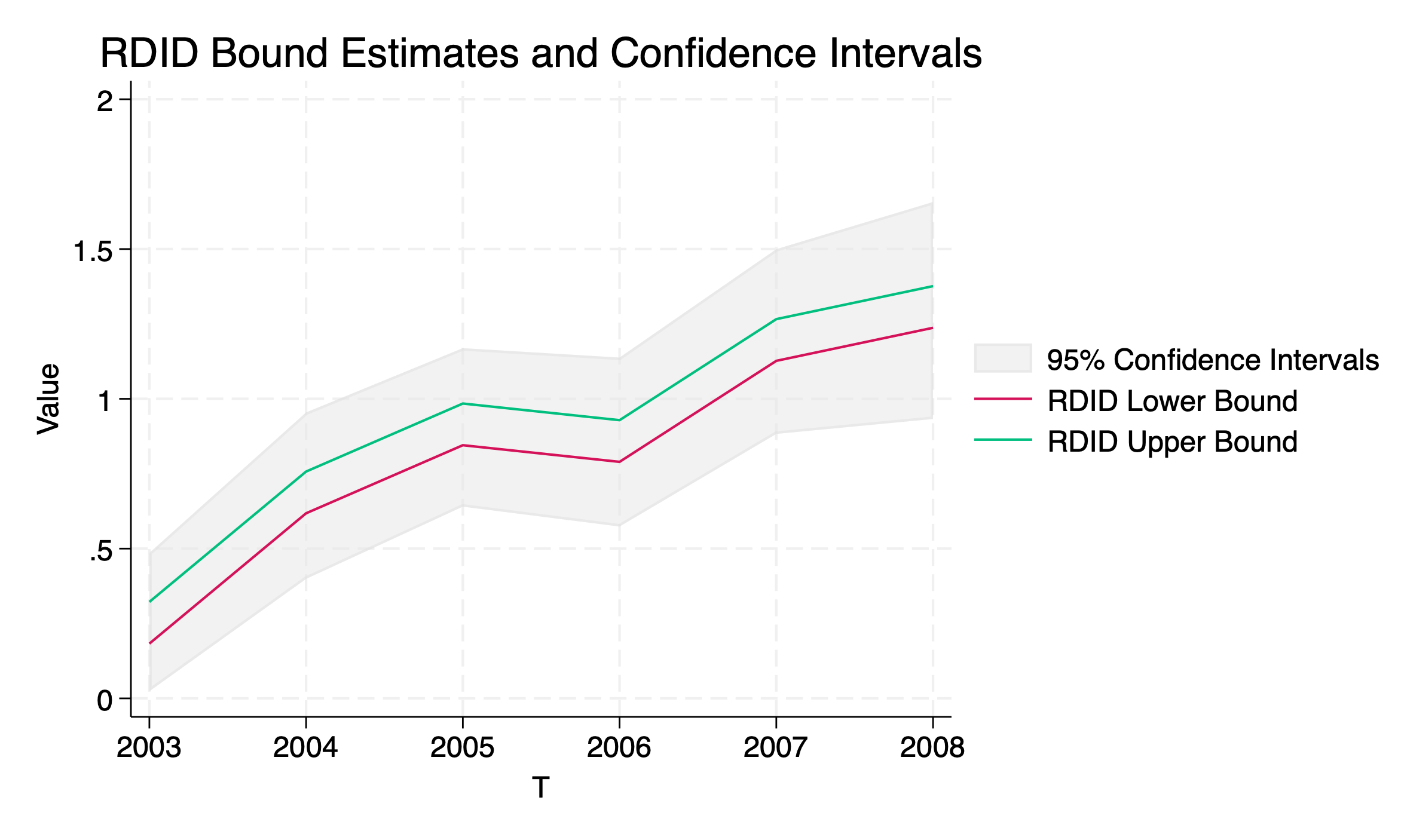}
	\caption{Figure produced by \texttt{rdid\_dy} with {\tt rdidtype(0)}}
	\label{fig.dy1}
\end{figure}

Recall that \texttt{rdid\_dy} command is a wrapper command that implements RDID estimation separately for each levels of a time variable in the post-treatment period and collects the results. 
In the following runs, the same options are specified as before, but the option {\tt tname(year)} is used to denote the variable for the post-treatment periods.
Also, the following run produces Figure \ref{fig.dy1} that summarizes the output.

\begin{stlog}
\input{logs/log_rdid_dy_0.log.tex}\nullskip
\end{stlog}

\begin{figure}[h]
	\centering
	\includegraphics[width=0.8\textwidth]{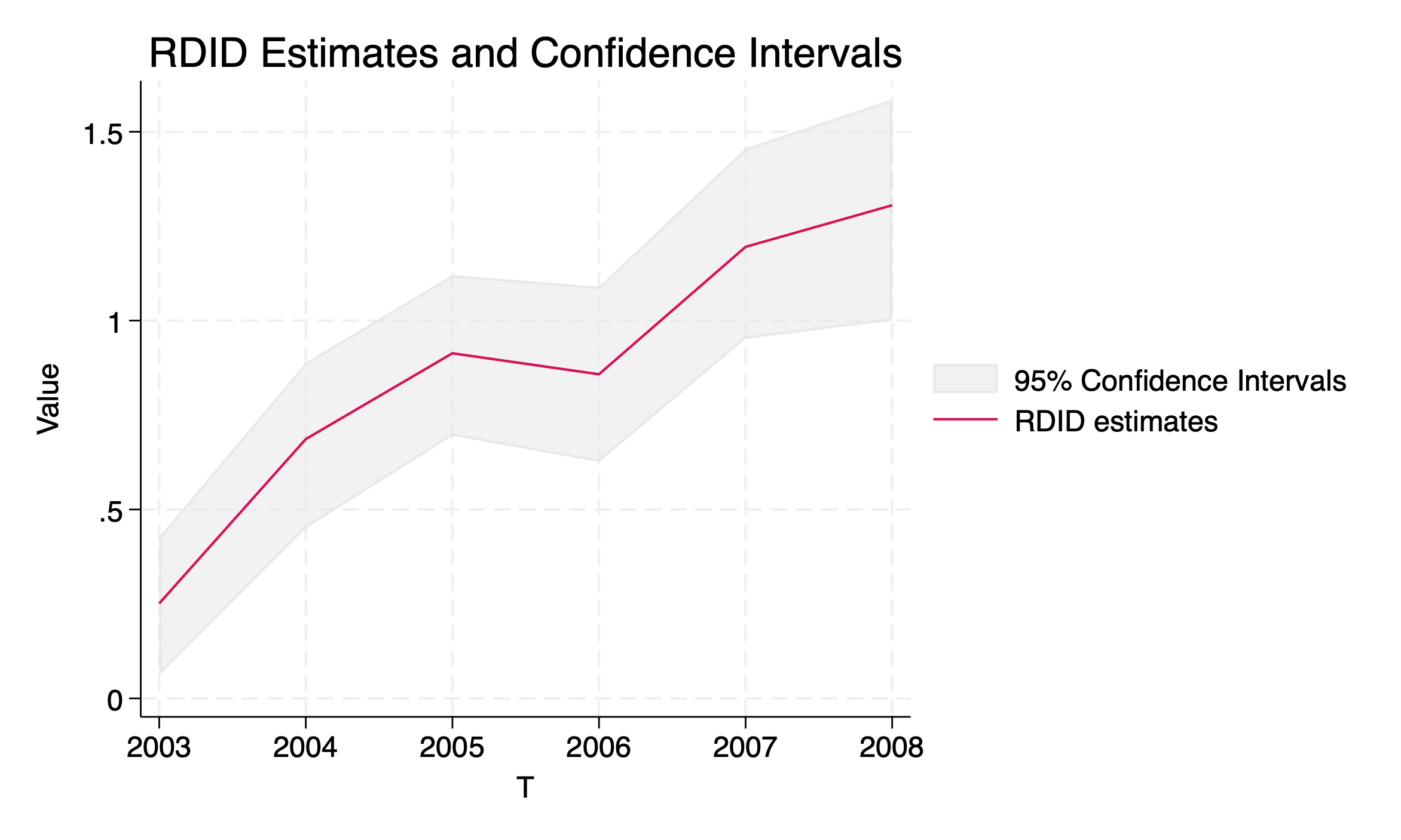}
	\caption{Figure produced by \texttt{rdid\_dy} with {\tt rdidtype(1)}}
	\label{fig.dy2}
\end{figure}

Note that \texttt{rdid\_dy} command can be used for the option {\tt rdidtype(1)} or {\tt rdidtype(2)}, and the following code runs for collecting PO-RDID with $\mathcal{L}_1$ (default), producing Figure \ref{fig.dy2} that summarizes the output.

\begin{stlog}
\input{logs/log_rdid_dy_1.log.tex}\nullskip
\end{stlog}

\clearpage

\subsection{Effects of minimum wage increases on teen employment}\label{sec.dube}

Following \cite{CallawaySantAnna2021} and \cite{dube2016}, we used the Quarterly Workforce Indicators (QWI) data to collect the first quarter teen employment as our outcome variable.
\cite{CallawaySantAnna2021} considered 7 years of periods between 2001 and 2007 where the federal minimum wage did not change over time, and 3 different control groups of $g=2004, 2006,$ and $2007$ with states that raised their minimum wage in or right before the beginning of years 2004, 2006, and 2007, respectively.
The specific timing of the raise can be found in \cite{CallawaySantAnna2021}, and it should be noted that there is some heterogeneity in the size of the minimum wage increase within each group.
The control group consists of states that did not raise their minimum wage during this period.

The following code specifies the option {\tt gname(G)} for the cohort index variable, which has four levels: 0, 2004, 2006, and 2007. 
Note that because the options {\tt postname(\varname)} and {\tt infoname(\varname)} are not specified, all periods from 2004 onward are treated as post-treatment periods, and all periods before 2004 are treated as informational elements.

\begin{stlog}
\input{logs/log_rdidstag.log.tex}\nullskip
\end{stlog}

\begin{figure}[h]
	\centering
	\includegraphics[width=0.8\textwidth]{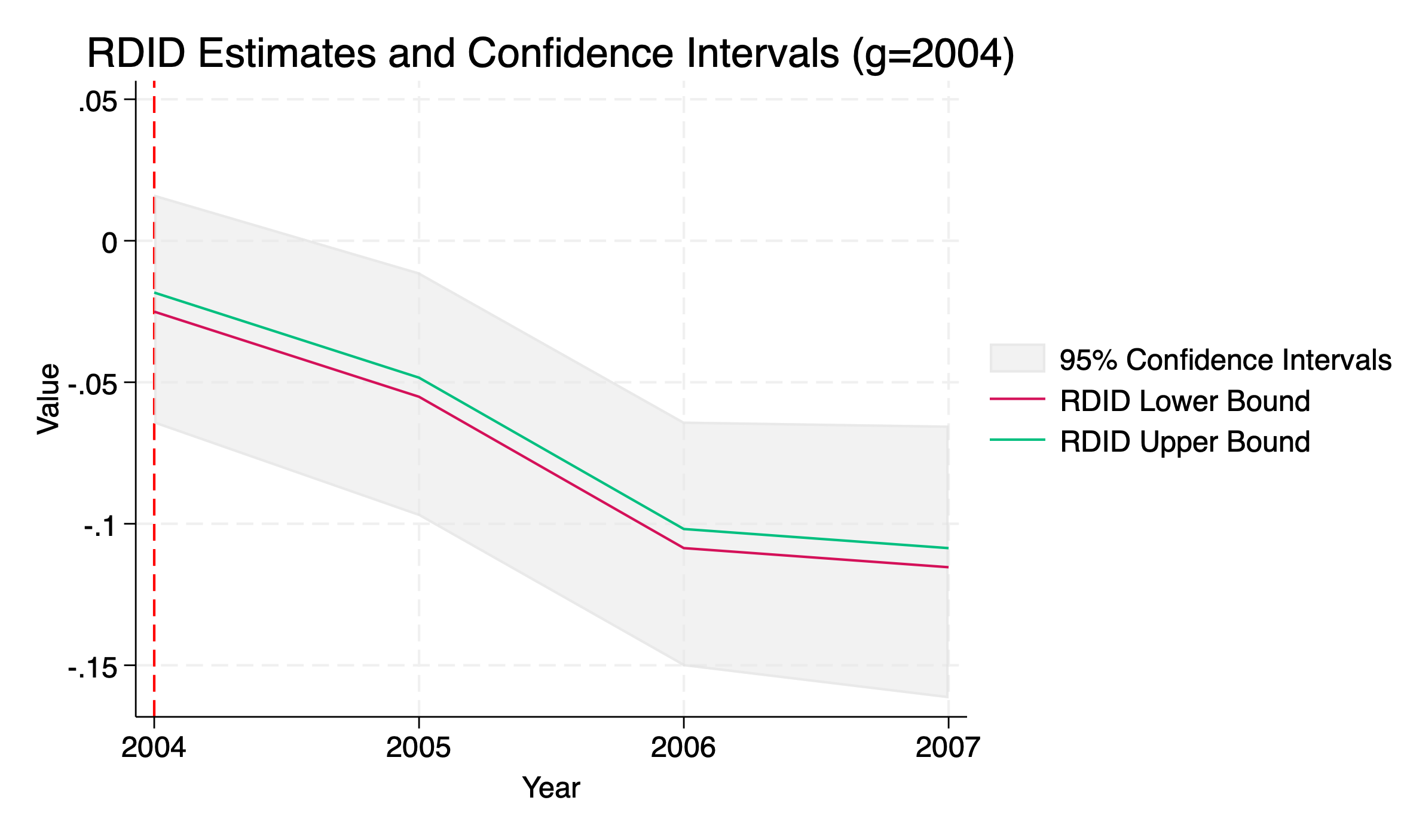}
	\caption{Figure produced by \texttt{rdidstag} ($g=2004$)}
\end{figure}

\begin{figure}[h]
	\centering
	\includegraphics[width=0.8\textwidth]{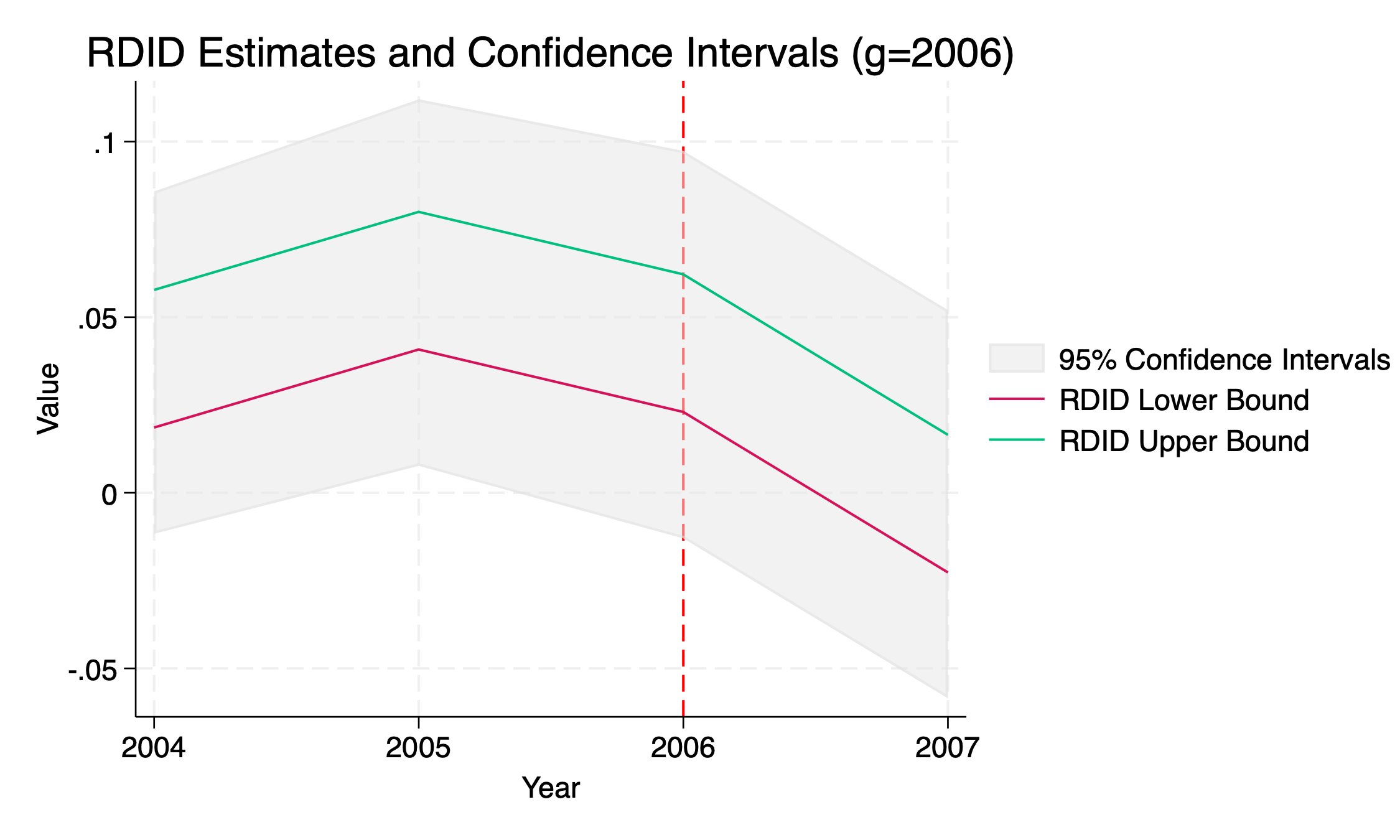}
	\caption{Figure produced by \texttt{rdidstag} ($g=2006$)}
\end{figure}

\begin{figure}[h]
	\centering
	\includegraphics[width=0.8\textwidth]{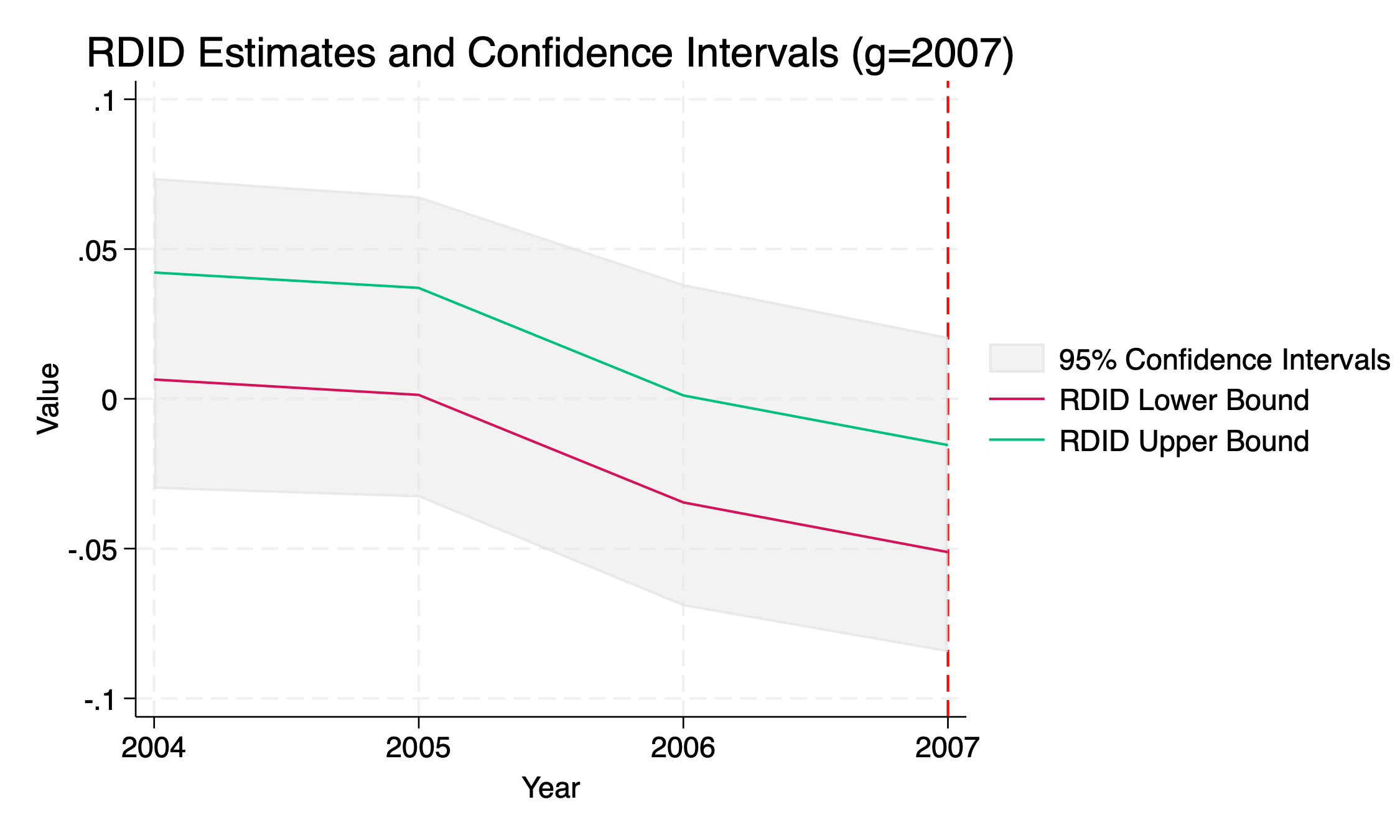}
	\caption{Figure produced by \texttt{rdidstag} ($g=2007$)}
\end{figure}

\clearpage 
\section{Monte-Carlo Simulations}

\subsection{Modified Example 2: \citeauthor{Ashenfelter1978}'s (\citeyear{Ashenfelter1978}) dip}

Consider 
		\begin{eqnarray*}
			\left\{ \begin{array}{lcl}
				Y_{it}&=&(1+\vert t \vert + t^2)U_{i}+\theta D_i*t\mathbbm{1}\{t\geq 0\} + 4\cdot\varepsilon_{it} \\
				D_i&=&\mathbbm{1}\{U_i\geq 1\}\\ 
				U_i &\sim& N(0,1)\\ 
				\varepsilon_{it} &\sim& N(0,1)
			\end{array} \right.
		\end{eqnarray*} for $t \in \mathcal T_0 \cup \{1\}$ where $\mathcal I_0=\mathcal T_0=\{-2,-1,0\}$.

In this model, $SB_t=(1+\vert t \vert + t^2)(\alpha_1-\alpha_0)$ where $\alpha_1=\frac{\phi(1)}{1-\Phi(1)}\approx 1.53$ and $\alpha_0=-\frac{\phi(1)}{\Phi(1)} \approx -0.29$. 
We have $SB_0= \alpha_1-\alpha_0 \neq 3(\alpha_1-\alpha_0) =SB_{-1} \neq SB_{-2}=7(\alpha_1-\alpha_0)$ and $SB_0=\alpha_1-\alpha_0 \neq 3(\alpha_1-\alpha_0)=SB_1$.  
So, the standard parallel trends assumption does not hold. 
However, the selection bias $SB_1$ in period 1 belongs to the convex hull of all selection biases in period 0, i.e., $SB_1 \in [\min\{SB_0,SB_{-1},SB_{-2}\}, \max\{SB_0,SB_{-1},SB_{-2}\}]=[\alpha_1-\alpha_0,7 (\alpha_1-\alpha_0)]$. Hence, our identifying assumption holds. 

We have $\theta_{OLS}=3(\alpha_1-\alpha_0)+\theta$, and $\Theta_I=[\theta-4(\alpha_1-\alpha_0),\theta+2(\alpha_1-\alpha_0)]$. The true $ATT=\theta$, and the DID estimand is $\theta_{DID}=\theta_{OLS}-SB_0=\theta+2(\alpha_1-\alpha_0)$. 
Thus, the DID estimand is upward biased and the bias is equal to $2(\alpha_1-\alpha_0)$. 
The bounds are generally informative about the magnitude of the average treatment effect on the treated (ATT) and can identify the sign of the ATT in some circumstances. 
For example, when the true ATT is equal to $-5$ or $9$, our bounds as well as the standard DID estimand identify the correct sign. 
On the other hand, when the true ATT is equal to $-1$, our bounds do not identify any sign, as they contain zero. 
But, the standard DID estimand identifies a wrong sign, it shows that the ATT is positive while it is actually negative. 

The following table shows a series of Monte-Carlo simulations for this DGP for three different confidence intervals (CIs): (1) \citeauthor{ye2023negative}'s (\citeyear{ye2023negative}) bootstrap CI for $\Theta_I$, (2) \citeauthor{ye2023negative}'s (\citeyear{ye2023negative}) bootstrap CI for ATT, (3) union bounds for $\Theta_I$..
The number of bootstrap replicates is 300, and the number of simulations is 1,000 for each case of $N \in \{200, 500, 1000, 5000, 10000, 50000\}$.

\begin{table}[htbp]
\centering
\begin{tabular}{ccccccc}\hline  \hline  
& \multicolumn{2}{c}{(1) \citeauthor{ye2023negative} for bounds} & \multicolumn{2}{c}{(2) \citeauthor{ye2023negative} for ATT} & \multicolumn{2}{c}{(3) Union bounds} \\
\cmidrule(lr){2-3} \cmidrule(lr){4-5} \cmidrule(lr){6-7} 
 & $CP_{inf}$  & Avg. Length  & $CP_{inf}$  & Avg. Length  & $CP_{inf}$  & Avg. Length  \\ \hline  
$N=200$  	&  0.9300 & 20.7121 & 0.9430 & 20.3115 & 0.9610 & 21.1112 \\ \hline 
$N=500$  	&  0.9540 & 17.1497 & 0.9560 & 16.8730 & 0.9620 & 17.2212 \\ \hline 
$N=1,000$ 	&  0.9490 & 15.1823 & 0.9490 & 14.9536 & 0.9510 & 15.1933 \\ \hline 
$N=5,000$ 	&  0.9510 & 12.8223 & 0.9470 & 12.6881 & 0.9510 & 12.8223 \\ \hline 
$N=10,000$ 	&  0.9680 & 12.2608 & 0.9610 & 12.1640 & 0.9680 & 12.2608 \\ \hline 
$N=50,000$ 	&  0.9640 & 11.4876 & 0.9550 & 11.4448 & 0.9640 & 11.4876 \\ \hline 
  \end{tabular}
\end{table}

\subsection{Example 1 from the article: with a covariate}

Consider 
\begin{eqnarray*}
\left\{ \begin{array}{lcl}
Y_t&=&(1+ 0.5^tX)U+\theta X D*t\mathbbm{1}\{t\geq 0\}  \\ 
D&=&\mathbbm{1}\{U\geq 1\}\\ 
U &\sim & N(0,1) \\
X &\sim & Bern(p),
\end{array} \right.
\end{eqnarray*} %
where $X \indep  U$ and $\mathcal I_0=\mathcal X=\{0,1 \}$.

We have $SB_0(x)=(1+x)(\alpha_1-\alpha_0)$, and $SB_1(x)=(1+0.5x)(\alpha_1-\alpha_0)$ where $\alpha_1=\frac{\phi(1)}{1-\Phi(1)}\approx 1.53$ and $\alpha_0=-\frac{\phi(1)}{\Phi(1)} \approx -0.29$. 
We have $SB_0(x) \in [\alpha_1-\alpha_0, 2(\alpha_1-\alpha_0)]$ and $SB_1(x) \in \left[\alpha_1-\alpha_0, 1.5(\alpha_1-\alpha_0)\right] \subseteq [\alpha_1-\alpha_0, 2(\alpha_1-\alpha_0)]\equiv \Delta_{SB(\mathcal X)}$. 
So, the standard parallel trends assumption does not hold as $SB_0(x)\neq SB_1(x)$. 
However, the selection bias $SB_1(x) $ in period 1 belongs to the convex hull of all selection biases in period 0, i.e., $SB_1(x) \in \Delta_{SB(\mathcal X)}$. 
Hence, our identifying assumption holds.
We have $\theta_{OLS}(x)=(1+ 0.5 x)(\alpha_1-\alpha_0)+\theta x$, and our new bounds $ \Theta_I $ are obtained as $ATT(x) \in [\theta x - (1-0.5 x)(\alpha_1-\alpha_0),\theta x + 0.5 x(\alpha_1-\alpha_0)]$. 
The actual conditional ATT function is $ATT(x)=\theta x$, but the standard conditional DID estimand is $\theta_{DID}(x)=\theta x - 0.5 x (\alpha_1 - \alpha_0)$.

Moreover, using the distribution of $X$, we have an identified set for ATT
$$
\Theta_I = \left[ \left(\frac{1}{2}p - 1 \right)(\alpha_1-\alpha_0) + p\theta, \frac{1}{2}(\alpha_1-\alpha_0) + p\theta \right]
$$
where $ATT = p\theta$.
The simulation is implemented using $p=0.5$, and the results are presented in the following table.

\begin{table}[htbp]
\centering
\begin{tabular}{ccccccc}\hline  \hline  
& \multicolumn{2}{c}{(1) \citeauthor{ye2023negative} for bounds} & \multicolumn{2}{c}{(2) \citeauthor{ye2023negative} for ATT} & \multicolumn{2}{c}{(3) Union bounds} \\
\cmidrule(lr){2-3} \cmidrule(lr){4-5} \cmidrule(lr){6-7} 
 & $CP_{inf}$  & Avg. Length  & $CP_{inf}$  & Avg. Length  & $CP_{inf}$  & Avg. Length  \\ \hline  
$N=200$  	&  0.9340 & 5.4190 & 0.9620 & 5.3556 & 0.9340 & 5.4190 \\ \hline   
$N=500$  	&  0.9550 & 4.0948 & 0.9710 & 4.0366 & 0.9550 & 4.0948 \\ \hline   
$N=1,000$ 	&  0.9480 & 3.4183 & 0.9620 & 3.3703 & 0.9480 & 3.4183 \\ \hline   
$N=5,000$ 	&  0.9500 & 2.5282 & 0.9560 & 2.4942 & 0.9500 & 2.5282 \\ \hline   
$N=10,000$ 	&  0.9520 & 2.3183 & 0.9600 & 2.2887 & 0.9520 & 2.3183 \\ \hline   
$N=50,000$ 	&  0.9610 & 2.0405 & 0.9530 & 2.0237 & 0.9610 & 2.0405 \\ \hline   
  \end{tabular}
\end{table}

\subsection{The modified staggered adoption design}

Consider
\begin{eqnarray*}
Y_{it} &=& (1+t^2)U_i + \varepsilon_{it} + \sum_{s=1}^T \theta_{is} D_{is} \mathbbm{1}\{s \leq t\} \quad \mathrm{ for } \,\, t=0, \ldots, T \\
D_{it} &=& \mathbbm{1}\bigg\{U_i \geq 2 - \frac{t}{T} \bigg\}
\end{eqnarray*} 
where $U_i \indep (\{\varepsilon_{it}, \theta_{it} \}_{t=1}^T )$, $\theta_{it} \sim \mathcal{N}(\frac{1+t^2}{2}, 1)$, $\mathcal{I}_0 = \mathcal T_0 = \{-T, \ldots, 0 \}$, $\varepsilon_{it} \sim \mathcal{N}(t^2, 1)$, $U_i \sim \mathcal{U}_{[0, 2]}$, and $T=4$.

Note that the PT is violated as
\begin{align*}
& \mathbbm{E}[Y_{it}(g')-Y_{i0}(g') \mid G_i=g] - \mathbbm{E}[Y_{it}(g')-Y_{i0}(g') \mid G_i=g'] \\
& \qquad = \mathbbm{E}[t^2U_i \mid G_i = g] -  \mathbbm{E}[t^2U_i \mid G_i = g'], \\
& \qquad \neq 0,
\end{align*}
for $t > 0$ when $g \neq g'$, especially with $g' = \infty$ for the never-treated cohorts being the control units.

However, note that
\begin{align*}
SB_t[g' \rightarrow g] &= \mathbbm{E}[ Y_{it} (g') \mid G_i = g] -  \mathbbm{E}[ Y_{it} (g') \mid G_i = g'] \\
&= \mathbbm{E}[ (1+t^2)U_i \mid G_i = g] - \mathbbm{E}[ (1+t^2)U_i \mid G_i = g'],
\end{align*}
and
\begin{align*}
SB_t[g' \rightarrow g] &\in \Delta_{SB(\mathcal T_0)}[g' \rightarrow g]\\
&= \bigg[ \mathbbm{E}[U_i \mid G_i = g] - \mathbbm{E}[U_i \mid G_i = g'], (1+T^2)\big(\mathbbm{E}[ U_i \mid G_i = g] - \mathbbm{E}[U_i \mid G_i = g']\big) \bigg],
\end{align*}
for all $t=1, \ldots, T$.

In the case of $g' = \infty$, we have
\begin{align*}
ATT(g, t) &= \mathbbm{E}[Y_{it}(g) - Y_{it}(\infty) \mid G_i = g] \\
&= \mathbbm{E}\bigg[\sum_{s=g}^T \theta_s \mathbbm{1}\{s \leq t\} \bigg],
\end{align*}
and
\begin{align*}
\theta_{DIM}^t[\infty \rightarrow g] &= \mathbbm{E}[Y_{it} \mid G_i = g] - \mathbbm{E}[Y_{it} \mid G_i = \infty] \\
&= ATT(g, t) +  SB_t[\infty \rightarrow g].
\end{align*}

Hence, we have
\begin{align*}
\Theta(g, t) 
&= \bigg[ ATT(g, t) + (t^2 - T^2)\mu(g), ATT(g, t) + t^2 \mu(g) \bigg],
\end{align*}
where $\mu(g) \equiv \mathbbm{E}[U_{i} \mid G_i = g] - \mathbbm{E}[U_{i} \mid G_i = \infty]$.

Under this setup, we implemented a series of Monte-Carlo simulations with various $N$ values, and the following tables summarize the coverage probability ($CP_{inf}$) and the average length of the 95\% confidence interval for each case.

\begin{table}[htbp]
\footnotesize
\centering
\caption{Results with $g=1$}
\label{tab:attgt1}
\begin{tabular}{ccccccccc}\hline  \hline  
& \multicolumn{2}{c}{(1) $ATT(g, 1)$} & \multicolumn{2}{c}{(2)  $ATT(g, 2)$} & \multicolumn{2}{c}{(3)  $ATT(g, 3)$}& \multicolumn{2}{c}{(4)  $ATT(g, 4)$} \\
\cmidrule(lr){2-3} \cmidrule(lr){4-5} \cmidrule(lr){6-7} \cmidrule(lr){8-9} 
$N$ & $CP_{inf}$  & Avg. Len.  & $CP_{inf}$  & Avg. Len.  & $CP_{inf}$  & Avg. Len.  & $CP_{inf}$  & Avg. Len.  \\ \hline     
$100$  	&  0.9670 & 24.9483 & 0.9690 & 25.2785 & 0.9770 & 25.8242 & 0.9770 & 26.7418 \\ \hline         
$500$ 	&  0.9660 & 23.3114 & 0.9770 & 23.4608 & 0.9810 & 23.6989 & 0.9810 & 24.0988 \\ \hline         
$1000$ 	&  0.9630 & 22.9296 & 0.9710 & 23.0335 & 0.9840 & 23.2027 & 0.9810 & 23.4838 \\ \hline         
$5000$ 	&  0.9720 & 22.4170 & 0.9810 & 22.4637 & 0.9840 & 22.5398 & 0.9880 & 22.6645 \\ \hline   
$10000$	&  0.9680 & 22.2932 & 0.9710 & 22.3263 & 0.9820 & 22.3799 & 0.9860 & 22.4681 \\ \hline   
  \end{tabular}
\end{table}

\begin{table}[htbp]
\footnotesize
\centering
\caption{Results with $g=2$}
\begin{tabular}{ccccccccc}\hline  \hline  
& \multicolumn{2}{c}{(1) $ATT(g, 1)$} & \multicolumn{2}{c}{(2)  $ATT(g, 2)$} & \multicolumn{2}{c}{(3)  $ATT(g, 3)$}& \multicolumn{2}{c}{(4)  $ATT(g, 4)$} \\
\cmidrule(lr){2-3} \cmidrule(lr){4-5} \cmidrule(lr){6-7} \cmidrule(lr){8-9} 
$N$ & $CP_{inf}$  & Avg. Len.  & $CP_{inf}$  & Avg. Len.  & $CP_{inf}$  & Avg. Len.  & $CP_{inf}$  & Avg. Len.  \\ \hline         
$100$  	&  0.9560 & 20.7114 & 0.9660 & 21.0820 & 0.9700 & 21.6513 & 0.9700 & 22.5811 \\ \hline           
$500$ 	&  0.9580 & 19.1992 & 0.9720 & 19.3615 & 0.9710 & 19.6176 & 0.9740 & 20.0279 \\ \hline           
$1000$ 	&  0.9620 & 18.8385 & 0.9700 & 18.9532 & 0.9690 & 19.1349 & 0.9850 & 19.4250 \\ \hline           
$5000$ 	&  0.9610 & 18.3819 & 0.9750 & 18.4335 & 0.9830 & 18.5146 & 0.9800 & 18.6438 \\ \hline 
$10000$	&  0.9700 & 18.2707 & 0.9720 & 18.3071 & 0.9880 & 18.3645 & 0.9830 & 18.4559 \\ \hline 
  \end{tabular}
\end{table}

\begin{table}[htbp]
\footnotesize
\centering
\caption{Results with $g=3$}
\begin{tabular}{ccccccccc}\hline  \hline  
& \multicolumn{2}{c}{(1) $ATT(g, 1)$} & \multicolumn{2}{c}{(2)  $ATT(g, 2)$} & \multicolumn{2}{c}{(3)  $ATT(g, 3)$}& \multicolumn{2}{c}{(4)  $ATT(g, 4)$} \\
\cmidrule(lr){2-3} \cmidrule(lr){4-5} \cmidrule(lr){6-7} \cmidrule(lr){8-9} 
$N$ & $CP_{inf}$  & Avg. Len.  & $CP_{inf}$  & Avg. Len.  & $CP_{inf}$  & Avg. Len.  & $CP_{inf}$  & Avg. Len.  \\ \hline  
$100$  	&  0.9670 & 16.6965 & 0.9760 & 16.8386 & 0.9680 & 17.4816 & 0.9860 & 18.4511 \\ \hline   
$500$ 	&  0.9620 & 15.2021 & 0.9780 & 15.2637 & 0.9800 & 15.5432 & 0.9800 & 15.9710 \\ \hline   
$1000$ 	&  0.9580 & 14.8394 & 0.9740 & 14.8832 & 0.9880 & 15.0802 & 0.9830 & 15.3824 \\ \hline   
$5000$ 	&  0.9630 & 14.3781 & 0.9760 & 14.3974 & 0.9790 & 14.4852 & 0.9800 & 14.6201 \\ \hline  
$10000$	&  0.9680 & 14.2669 & 0.9730 & 14.2806 & 0.9760 & 14.3425 & 0.9770 & 14.4382 \\ \hline  
  \end{tabular}
\end{table}

\begin{table}[htbp]
\footnotesize
\centering
\caption{Results with $g=4$}
\begin{tabular}{ccccccccc}\hline  \hline  
& \multicolumn{2}{c}{(1) $ATT(g, 1)$} & \multicolumn{2}{c}{(2)  $ATT(g, 2)$} & \multicolumn{2}{c}{(3)  $ATT(g, 3)$}& \multicolumn{2}{c}{(4)  $ATT(g, 4)$} \\
\cmidrule(lr){2-3} \cmidrule(lr){4-5} \cmidrule(lr){6-7} \cmidrule(lr){8-9} 
$N$ & $CP_{inf}$  & Avg. Len.  & $CP_{inf}$  & Avg. Len.  & $CP_{inf}$  & Avg. Len.  & $CP_{inf}$  & Avg. Len.  \\ \hline   
$100$  	& 0.9690 & 12.7222 & 0.9820 & 12.8685 & 0.9820 & 13.3045 & 0.9830 & 14.3359 \\ \hline  
$500$ 	& 0.9590 & 11.2131 & 0.9740 & 11.2757 & 0.9820 & 11.4668 & 0.9800 & 11.9167 \\ \hline  
$1000$ 	& 0.9650 & 10.8464 & 0.9710 & 10.8904 & 0.9790 & 11.0256 & 0.9780 & 11.3440 \\ \hline  
$5000$ 	& 0.9620 & 10.3813 & 0.9740 & 10.4005 & 0.9780 & 10.4611 & 0.9760 & 10.6031 \\ \hline 
$10000$	& 0.9640 & 10.2689 & 0.9660 & 10.2825 & 0.9800 & 10.3252 & 0.9770 & 10.4256 \\ \hline 
  \end{tabular}
\end{table}

\clearpage

\section{Conclusion}

In this article, we introduce the \texttt{rdid}, \texttt{rdid\_dy}, and \texttt{rdidstag} commands to implement the RDID framework by \cite{bk2023rdid} for both the canonical $2 \times 2$ DID settings and staggered adoption designs. 
By allowing users to specify their information set as pre-treatment periods, the commands enhance robustness to potential violations of the parallel trends assumption in both settings.
We anticipate that continued enhancements and community feedback will further increase its effectiveness and adoption in empirical research.

\section{Programs and supplemental material}

To obtain the latest stable versions of \texttt{rdid}, \texttt{rdid\_dy}, and \texttt{rdidstag} from our website, refer to the installation instructions at \href{https://github.com/KyunghoonBan/rdid}{https://github.com/KyunghoonBan/rdid}.

\bibliographystyle{sj}
\bibliography{mybib}

\clearpage
\end{document}

%% file: logs/log_rdid_0.log.tex
. use "analysis.dta"
{\smallskip}
. drop if sector != 1
(20,519 observations deleted)
{\smallskip}
. rdid area_tob hhsize educ_scale age, treat(treatment) post(policy2) ///
>         info(year)  b(1000) cl(hhno)
 **** RDID Estimation version 1.8 **** 
Y name: area_tob
D name: treatment
X name(s): hhsize educ_scale age
{\smallskip}
\HLI{16}{\TOPT}\HLI{19}
    Y: area_tob {\VBAR}       LB        UB 
\HLI{16}{\PLUS}\HLI{19}
           RDID {\VBAR}   0.8086    0.9479 
           CI_1 {\VBAR}   0.6203    1.1257 
           CI_2 {\VBAR}   0.6235    1.1222 
           CI_3 {\VBAR}   0.6187    1.1257 
\HLI{16}{\BOTT}\HLI{19}
* RDID: Point estimates for RDID bounds
* CI_1: Confidence interval for the bounds (Ye et al.)
* CI_2: Confidence interval for the ATT (Ye et al.)
* CI_3: Confidence interval for the bounds (Union Bounds)

%% file: logs/log_rdid_1.log.tex
. rdid area_tob hhsize educ_scale age, treat(treatment) post(policy2) ///
>         info(year) rdidtype(1) b(1000) cl(hhno)
 **** RDID Estimation version 1.8 **** 
Y name: area_tob
D name: treatment
X name(s): hhsize educ_scale age
{\smallskip}
\HLI{16}{\TOPT}\HLI{23}
    Y: area_tob {\VBAR}     PE   CI_LB   CI_UB 
\HLI{16}{\PLUS}\HLI{23}
             L1 {\VBAR} 0.8770  0.6732  1.0806 
             L2 {\VBAR} 0.8778  0.6911  1.0753 
           Linf {\VBAR} 0.8783  0.6978  1.0710 
\HLI{16}{\BOTT}\HLI{23}

%% file: logs/log_rdid_2.log.tex
. rdid area_tob hhsize educ_scale age, treat(treatment) post(policy2) ///
>         info(year) rdidtype(2) peval(2005.5) b(1000) cl(hhno)
 **** RDID Estimation version 1.8 **** 
Y name: area_tob
D name: treatment
X name(s): hhsize educ_scale age
{\smallskip}
\HLI{16}{\TOPT}\HLI{39}
                {\VBAR}     PE   CI_LB   CI_UB  TAU_DR  SB_hat 
\HLI{16}{\PLUS}\HLI{39}
Y               {\VBAR}                                       
       area_tob {\VBAR} 0.7239  0.4760  0.9725  1.9154  1.1914 
\HLI{16}{\BOTT}\HLI{39}

%% file: logs/log_rdid_dy_0.log.tex
. use "analysis.dta"
{\smallskip}
. drop if sector != 1
(20,519 observations deleted)
{\smallskip}
. rdid_dy area_tob hhsize educ_scale age, treat(treatment) post(policy2) ///
>         info(year) tname(year) fig b(1000) cl(hhno)
file{\bftt{ rdid_dy.png}} saved as PNG format
{\smallskip}
\HLI{16}{\TOPT}\HLI{39}
        T: year {\VBAR}  RDID_LB   RDID_UB     CI_LB     CI_UB 
\HLI{16}{\PLUS}\HLI{39}
           2003 {\VBAR}   0.1829    0.3221    0.0239    0.4855 
           2004 {\VBAR}   0.6181    0.7573    0.3988    0.9545 
           2005 {\VBAR}   0.8452    0.9844    0.6401    1.1692 
           2006 {\VBAR}   0.7897    0.9289    0.5738    1.1377 
           2007 {\VBAR}   1.1269    1.2661    0.8829    1.4996 
           2008 {\VBAR}   1.2372    1.3764    0.9328    1.6571 
\HLI{16}{\BOTT}\HLI{39}
* Confidence intervals are obtained for the bounds (Ye et al.)

%% file: logs/log_rdid_dy_1.log.tex
. rdid_dy area_tob hhsize educ_scale age, rdidtype(1) treat(treatment) ///
>         post(policy2) info(year) tname(year) fig b(1000) cl(hhno)
file{\bftt{ rdid_dy.png}} saved as PNG format
{\smallskip}
\HLI{16}{\TOPT}\HLI{29}
        T: year {\VBAR}  RDID_PE     CI_LB     CI_UB 
\HLI{16}{\PLUS}\HLI{29}
           2003 {\VBAR}   0.2513    0.0605    0.4273 
           2004 {\VBAR}   0.6865    0.4497    0.8893 
           2005 {\VBAR}   0.9136    0.6951    1.1206 
           2006 {\VBAR}   0.8581    0.6253    1.0898 
           2007 {\VBAR}   1.1953    0.9517    1.4561 
           2008 {\VBAR}   1.3056    0.9999    1.5860 
\HLI{16}{\BOTT}\HLI{29}
* RDID estimates are obtained as PO-RDID estimates (L1)

%% file: logs/log_rdidstag.log.tex
. use "qwi_minwageclean2.dta"
{\smallskip}
. destring years, replace
years: all characters numeric; replaced as int
{\smallskip}
. rdidstag lnemp_0A01_BS lnpop lnteenpop lnemp_0A00_BS, gname(G) tname(years) cl(countyfips) fig
 **** RDID Estimation version 1.7 **** 
Y name: lnemp_0A01_BS
G name: G
X name(s): lnpop lnteenpop lnemp_0A00_BS
postname is not specified: using units with T >= min(G) as post-period units.
infoname is not specified: using pre-period T < min(G) as the information set.
{\smallskip}
\HLI{16}{\TOPT}\HLI{39}
       ATT(G/T) {\VBAR}  RDID_LB   RDID_UB   95CI_LB   95CI_UB 
\HLI{16}{\PLUS}\HLI{39}
 ATT(2004/2004) {\VBAR}  -0.0250   -0.0183   -0.0646    0.0164 
 ATT(2004/2005) {\VBAR}  -0.0551   -0.0484   -0.0973   -0.0111 
 ATT(2004/2006) {\VBAR}  -0.1086   -0.1018   -0.1504   -0.0639 
 ATT(2004/2007) {\VBAR}  -0.1153   -0.1086   -0.1617   -0.0653 
 ATT(2006/2004) {\VBAR}   0.0186    0.0578   -0.0117    0.0858 
 ATT(2006/2005) {\VBAR}   0.0408    0.0800    0.0077    0.1121 
 ATT(2006/2006) {\VBAR}   0.0230    0.0622   -0.0130    0.0974 
 ATT(2006/2007) {\VBAR}  -0.0227    0.0165   -0.0585    0.0520 
 ATT(2007/2004) {\VBAR}   0.0065    0.0422   -0.0301    0.0738 
 ATT(2007/2005) {\VBAR}   0.0013    0.0370   -0.0329    0.0676 
 ATT(2007/2006) {\VBAR}  -0.0346    0.0012   -0.0692    0.0383 
 ATT(2007/2007) {\VBAR}  -0.0512   -0.0154   -0.0846    0.0207 
\HLI{16}{\BOTT}\HLI{39}
- Information elements: 2001 2002 2003
- Post-periods: 2004 2005 2006 2007
- Groups: 2004 2006 2007
file{\bftt{ rdidstag_g2004.png}} saved as PNG format
file{\bftt{ rdidstag_g2006.png}} saved as PNG format
file{\bftt{ rdidstag_g2007.png}} saved as PNG format

%% file: rdid_10062024.bbl
\ifnum 16=1 \def\bibname{Reference}
\else \def\bibname{References} \fi
\begin{thebibliography}{16}
\expandafter\ifx\csname natexlab\endcsname\relax\def\natexlab#1{#1}\fi
\expandafter\ifx\csname url\endcsname\relax
  \def\url#1{\texttt{#1}}\fi
\expandafter\ifx\csname urlprefix\endcsname\relax\def\urlprefix{URL }\fi

\bibitem[{Ashenfelter(1978)}]{Ashenfelter1978}
Ashenfelter, O. 1978.
\newblock Estimating the Effect of Training Programs on Earnings.
\newblock \emph{The Review of Economics and Statistics} 60(1): 47--57.

\bibitem[{Ban and Kédagni(2023)}]{bk2023rdid}
Ban, K., and D.~Kédagni. 2023.
\newblock Robust Difference-in-differences Models.
\urlprefix\url{https://arxiv.org/abs/2211.06710}.
\bibitem[{Bravo et~al.\@(2022)Bravo, Roth, and Rambachan}]{bravo2022honestdid}
Bravo, M.~C., J.~Roth, and A.~Rambachan. 2022.
\newblock Honestdid: Stata module implementing the honestdid r package .

\bibitem[{Cai(2016)}]{cai2016}
Cai, J. 2016.
\newblock {The Impact of Insurance Provision on Household Production and
  Financial Decisions}.
\newblock \emph{American Economic Journal: Economic Policy} 8(2): 44--88.

\bibitem[{Callaway and Sant’Anna(2021)}]{CallawaySantAnna2021}
Callaway, B., and P.~H. Sant’Anna. 2021.
\newblock {Difference-in-Differences with multiple time periods}.
\newblock \emph{Journal of Econometrics} 225(2): 200--230.

\bibitem[{de~Chaisemartin et~al.\@(2019)de~Chaisemartin, D’haultf{\oe}uille,
  and Guyonvarch}]{de2019fuzzy}
de~Chaisemartin, C., X.~D’haultf{\oe}uille, and Y.~Guyonvarch. 2019.
\newblock Fuzzy differences-in-differences with stata.
\newblock \emph{The Stata Journal} 19(2): 435--458.

\bibitem[{Dube et~al.\@(2016)Dube, Lester, and Reich}]{dube2016}
Dube, A., T.~W. Lester, and M.~Reich. 2016.
\newblock {Minimum Wage Shocks, Employment Flows, and Labor Market Frictions}.
\newblock \emph{Journal of Labor Economics} 34(3): 663--704.

\bibitem[{Han(2021)}]{Han2021}
Han, S. 2021.
\newblock Identification in nonparametric models for dynamic treatment effects.
\newblock \emph{Journal of Econometrics} 225: 132--147.

\bibitem[{Houngbedji(2016)}]{houngbedji2016abadie}
Houngbedji, K. 2016.
\newblock Abadie's semiparametric difference-in-differences estimator.
\newblock \emph{The Stata Journal} 16(2): 482--490.

\bibitem[{Mora and Reggio(2015)}]{mora2015didq}
Mora, R., and I.~Reggio. 2015.
\newblock didq: A command for treatment-effect estimation under alternative
  assumptions.
\newblock \emph{The Stata Journal} 15(3): 796--808.

\bibitem[{Rios-Avila et~al.\@(2023)Rios-Avila, Sant'Anna, and
  Callaway}]{rios2023csdid}
Rios-Avila, F., P.~Sant'Anna, and B.~Callaway. 2023.
\newblock CSDID: Stata module for the estimation of Difference-in-Difference
  models with multiple time periods .

\bibitem[{Rios-Avila et~al.\@(2022)Rios-Avila, Sant'Anna, and
  Naqvi}]{rios2022drdid}
Rios-Avila, F., P.~Sant'Anna, and A.~Naqvi. 2022.
\newblock DRDID: Stata module for the estimation of doubly robust
  difference-in-difference models .

\bibitem[{Robins(1986)}]{Robins1986}
Robins, J.~M. 1986.
\newblock A new approach to causal inference in mortality studies with a
  sustained exposure period--application to control of the healthy worker
  survivor effect.
\newblock \emph{Math. Modelling} 7: 1393--1512.

\bibitem[{Robins(1987)}]{Robins1987}
\mbox{\vrule width30.25006ptheight2.62222ptdepth-2.25222pt}. 1987.
\newblock Addendum to ``A new approach to causal inference in mortality studies
  with a sustained exposure period --- application to control of the healthy
  worker survivor effect''.
\newblock \emph{Comput. Math. Appl.} 14(9--12): 923--945.

\bibitem[{Villa(2016)}]{villa2016diff}
Villa, J.~M. 2016.
\newblock diff: Simplifying the estimation of difference-in-differences
  treatment effects.
\newblock \emph{The Stata Journal} 16(1): 52--71.

\bibitem[{Ye et~al.\@(2023)Ye, Keele, Hasegawa, and Small}]{ye2023negative}
Ye, T., L.~Keele, R.~Hasegawa, and D.~S. Small. 2023.
\newblock {A Negative Correlation Strategy for Bracketing in
  Difference-in-Differences}.
\newblock \emph{Journal of the American Statistical Association}
  ahead-of-print(ahead-of-print): 1--13.

\end{thebibliography}
